\documentclass{v23windows}
\usepackage{mathtools,amsmath,empheq,slashed}
\usepackage{amssymb}

\def\be{\begin{equation}}
\def\ee{\end{equation}}
\def\bea{\begin{eqnarray}}
\def\eea{\end{eqnarray}}

\newcommand{\eq}[1]{Eq.~(\ref{#1})}
\newcommand{\bib}[1]{Ref.~\cite{#1}}

\newcommand{\fig}[1]{Fig.~\ref{#1}}

\begin{document}
\vspace*{4cm}
\title{Gravitational positivity in electroweak sector}

\author{Tran Quang Loc$^{a,b}$}


\address{
$^{a}$Department of Theoretical Physics, University of Science, Ho Chi Minh City 70000, Vietnam\\
$^{b}$Vietnam National University, Ho Chi Minh City 70000, Vietnam
}

\maketitle\abstracts{
This study investigates the compatibility of the electroweak sector of particle physics with quantum gravity, under the assumption that the conventional S-matrix positivity bounds can be extended to gravitational context. It focuses on constraints implied by these bounds to the weak couplings of the Weinberg-Salam model coupled to gravity, analyzed through forward-limit of various $2\rightarrow 2$ scatterings, including $HH \rightarrow HH, H\gamma \rightarrow H\gamma, \gamma\gamma\rightarrow\gamma\gamma$. These constraints suggest possible extensions to the magnetic Weak Gravity Conjecture, relating gauge couplings with the EFT cutoff scale.
}

\section{Introduction}
The fundamental properties of the S-matrix—unitarity, analyticity, and crossing symmetry—lead to dispersion relations for forward elastic scattering amplitudes. This results in positivity bounds in the infrared (IR) spectrum, which imposes non-trivial constraints on the coefficients in effective field theories (EFTs) in low-energy amplitude calculations.

Infrared consistency of photon-graviton effective theory has been used to imply analog of Weak Gravity Conjecture (WGC) criterion in \bib{Cheung:2014ega}. More recently, gravitational positivity bounds in theories with a massless graviton can provide S-matrix evidence of the mild form of the WGC via Einstein-Maxwell low-energy EFT in \bib{Hamada:2018dde} and \bib{Bellazzini:2019xts} of the form
\begin{equation}
\sqrt{2}|Q| / M>1/M_{\mathrm{Pl}},
\end{equation}
where Q is the charge of the Abelian U(1) gauge theory coupled to gravity, M is the mass of extremal Black Holes, and $M_\text{Pl}$ represents Planck scale. 
An examination within the Standard Model (SM) minimally coupled to gravity also produces WGC-liked constraints on electron Yukawa couplings $y_e$ and the Weinberg angles $\theta_W$ in \bib{Aoki:2021ckh}
\begin{equation}
y_e \sin \theta_{\mathrm{W}}\geq\sqrt{\frac{11}{1440}} \frac{\Lambda}{M_{\mathrm{Pl}}}.
\end{equation}

Our study explores the aspect of positivity in the Weinberg-Salam (electroweak) model minimally coupled with gravity. Assuming a negligible Higgs mass and substantial fermion mass, we focus on constraints that positivity imposes on gauge couplings concerning the cut-off scale. The result suggests a possible extension to the magnetic WGC in \bib{Arkani-Hamed:2006emk} of the form \begin{equation}
g\gtrsim {\Lambda}/{M_{\mathrm{Pl}}}.
\end{equation}

\section{Gravitational Positivity Bounds}
\begin{figure}[!htb]\centering
\includegraphics[width=0.3\linewidth]{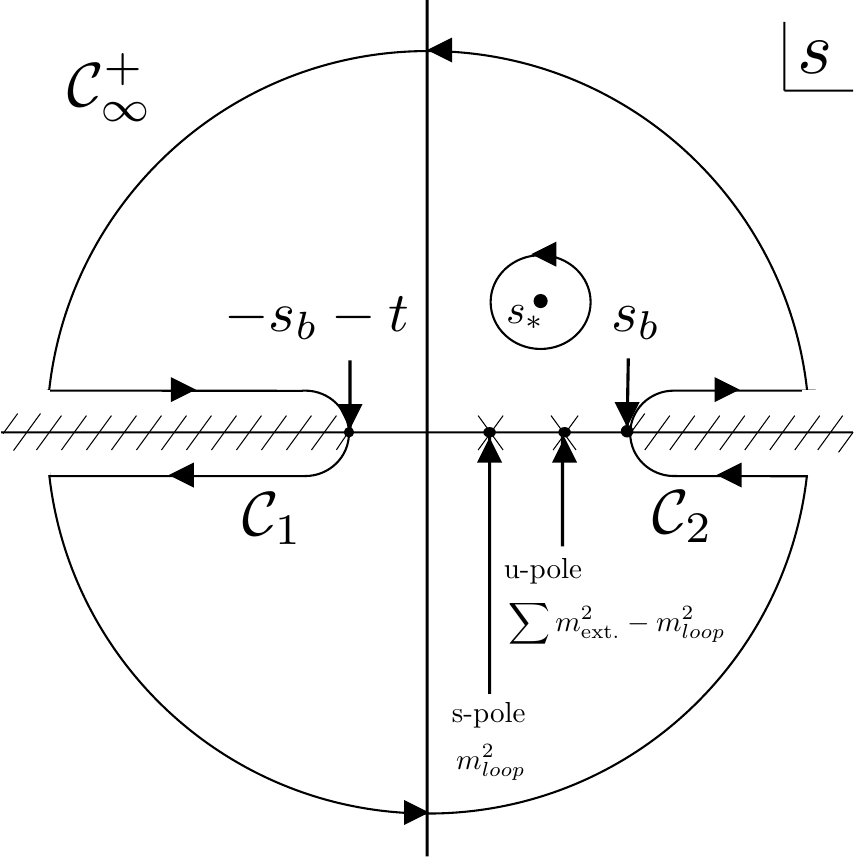}
\caption{Analytic structure of s-plane. The sum of Mandelstam variable equal total masses of external particles $s+t+u=\sum m_\text{ext.}^2$. At the reference point $s_*=\sum m_{\mathrm{ext.}}^2/2-t / 2+i \mu$, where $\mu>0$, the substracted amplitude is expressed through integrals along the branch cuts $\left(\mathcal{C}_1+\mathcal{C}_2\right)$ and the infinitely large semi-circles $\left(\mathcal{C}_{\infty}^{+}+\mathcal{C}_{\infty}^{-}\right)$. The contours $\mathcal{C}_1$ and $\mathcal{C}_2$ are straight lines defined by $\mathcal{C}_1:=$ $\left\{s^{\prime} \mid-\infty<s^{\prime}<-s_b-t\right\}$ with $s_b=4m_\text{loop}^2-\sum m_\text{ext.}^2$, and $\mathcal{C}_2:=\left\{s^{\prime} \mid 4m^2_\text{loop}<s^{\prime}<\infty\right\}$, respectively where $m_\text{loop}$ denote masses of particles inside loops.}
  \label{Analytic_plane}
\end{figure}
In the complex s-plane, with fixed $t < 0$, we posit the scattering amplitude $\mathcal{M}(s,t)$ is holomorphic in $s$ and exhibits $s \leftrightarrow u$ crossing symmetry, except for poles and cuts along the real axis. We further assume a mild ultraviolet (UV) behavior in the Regge limit as $\lim_{|s|\rightarrow\infty}\left|\frac{M(s,t < 0)}{s^2}\right|=0$. By subtracting the poles associated with $s$ and $u$ and denoting the modified amplitude as $\widetilde{\mathcal{M}}(s, t):=\mathcal{M}(s, t)-(s, u \text {-poles})$, we proceed to establish a dispersion relation considering the integration contour in Fig. \ref{Analytic_plane}. We further remove those known contributions below the cut off scale $\Lambda$ where the perturbation theory is valid to improve the dispersion relation:
	\begin{align}
	B^{(2)}(\Lambda, t)=\frac{8}{M_{\mathrm{Pl}}^2 t}+\frac{4}{\pi} \int_{\Lambda^2}^{\infty} \mathrm{d} s^{\prime} \frac{\operatorname{Im} \widetilde{\mathcal{M}}\left(s^{\prime}+i \epsilon, t\right)}{\left(s^{\prime}-s_*\right)^3} \label{Br2}.
\end{align}
The forward limit of Eq. \ref{Br2} presents a subtlety in the presence of gravity, primarily due to the singular first term on the right-hand side (RHS). This term in the forward limit depends on details of quantum gravity. Achieving a finite expression in the forward limit necessitates recognizing the cancellation of singular terms on both sides of equation  \ref{Br2} and careful evaluation of the $\mathcal{O}(t^0)$ term, e.g. Reggeization of graviton exchange, which has been explicitly detailed in \cite{Tokuda:2020mlf}.
Consequently, we can derive a sum rule that relates the infrared observable $B^{(2)}(\Lambda):=B^{(2)}(\Lambda, 0)$ to the properties of the S-matrix of quantum gravity. However, due to the lack of knowledge of the latter,
there is currently no proof of an inequality $B^{(2)}(\Lambda
)\geq0$ in contrast to the case for non-gravitational theories.

Interestingly, the positivity of $B^{(2)}(\Lambda)$ is connected to various versions of the Weak Gravity Conjecture (WGC) mentioned in the introduction. Therefore, we aim to investigate similar phenomena within a more comprehensive framework, specifically the Weinberg-Salam (electroweak) setup.
\section{Bounds on electroweak coupled to gravity}
In this section, we derive the twice-subtracted dispersion relation for three distinct 2-by-2 scattering processes: $\gamma \gamma \rightarrow \gamma \gamma, H\gamma \rightarrow H\gamma, HH \rightarrow HH$, where particles 1 and 2 are incoming, and 3 and 4 are outgoing, with $\pm$ signs indicating photon helicities. The momenta and helicities are defined in the Center of Mass (C.O.M.) frame, as outlined in \bib{Alberte:2020bdz}. We have utilized the incoming-outgoing convention, considering the summation over four helicity configurations denoted as $(h_1, h_2, h_3, h_4) \in \{(\pm, \pm, \pm, \pm), (\pm, \mp, \pm, \mp)\}$ to manifest $s\leftrightarrow u$ crossing for the $\gamma\gamma\rightarrow\gamma\gamma$ process. Additionally, for the $H\gamma\rightarrow H\gamma$ process, we have specifically taken $(h_1, h_3) = (+, +)$. We analyze electroweak (EW) minimally coupled to General Relativity (GR) amplitudes at $s\ll M_\text{Pl}^2$, considering EW and GR regimes separately. In each regime, we take the sum of the results from the QED, Weak, and Higgs sectors, 
\begin{align}
\mathcal{M}(s, t)=\mathcal{M}_{\mathrm{EW}}(s,t)+\mathcal{M}_{\mathrm{GR}}(s,t)=\sum_{i = \text{QED, Weak, Higgs}}\mathcal{M}_{\mathrm{EW,i}}(s,t)+\mathcal{M}_{\mathrm{GR,i}}(s,t).
\end{align}
\begin{figure}[h!]\centering
\includegraphics[width=0.95\linewidth]{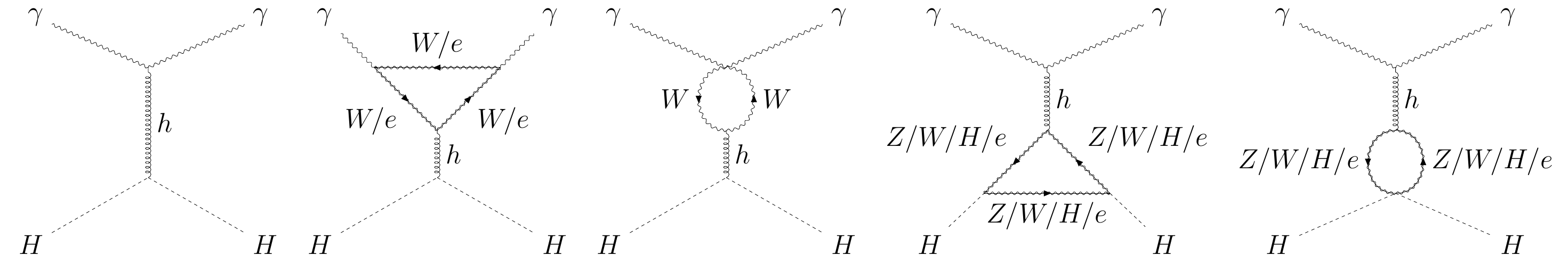}
\caption{Diagrams of dominant contribution for $H\gamma\rightarrow H\gamma$ process in GR sector}
  \label{loop-diagram}
\end{figure}
An example of diagrams of leading contribution in $H\gamma\rightarrow H\gamma$ process in GR regime is illustrated in Fig. \ref{loop-diagram}. With the forward scattering at high energy denoted as $\mathcal{A}(s):=\widetilde{\mathcal{M}}(s,0)$, the dispersion relation in each regime are derived by,
\begin{align}
B_{\mathrm{EW}}^{(2)}=\frac{4}{\pi} \int_{\Lambda^2}^{\infty} \frac{\operatorname{Im} \mathcal{A}(s)}{(s-s_*)^3} \mathrm{~d} s,\;\;\;\;\;\; B_{\mathrm{GR}}^{(2)}=\left.\frac{d^2}{d s^2} A(s)\right|_{s=s_*}-\frac{4}{\pi} \int_{s_{b}^2}^{\Lambda^2} \frac{\operatorname{Im} \mathcal{A}(s)}{(s-s_*)^3} \mathrm{~d} s .
\end{align}
In the EW sector, the process $HH \rightarrow HH$ the amplitude is predominantly governed by the Weak (W and Z)-boson loop. For the process  $H\gamma \rightarrow H\gamma$ and  $\gamma\gamma \rightarrow \gamma\gamma$, the dynamics involve only the W-boson and fermion loops, with the W-boson loop amplitude being dominant.

In the GR regime, the tree level diagrams' (pole) contribution is offset by the high energy integral of RHS of Eq. \ref{Br2}. Consequently, the one-loop becomes the leading contribution. The $\gamma\gamma \rightarrow \gamma\gamma$ process involves only the W-boson and fermion loops. For both the $H\gamma \rightarrow H\gamma$ and $HH \rightarrow HH$ processes in GR, there are contributions from pure fermion, Weak-boson, and Higgs loops. Additionally, the $HH \rightarrow HH$ process includes a unique loop structure formed by Higgs and one line of graviton. It is important to note that the s,u-diagrams arising from this loop structure are subject to infrared (IR) divergence due to the massless nature of the graviton. This problem is addressed by introducing a fictitious graviton mass, assuming the IR divergence is compensated by a soft graviton cloud \bib{Noumi:2021uuv}. Under this assumption, these diagrams lead to a subleading contribution that is suppressed by the factors $\Lambda^{-2} M_{\mathrm{Pl}}^{-2}$. More generally, in the $HH \rightarrow HH$ process for all loop types,
dominant contributions come from the t-channel. Our results are presented using the Planck scale \( M_{\text{Pl}} \) and the cutoff scale \( \Lambda \), along with five parameters: the Higgs mass \( m_H \), the Higgs field vacuum expectation value \( v \), the electron mass \( m_e \), and the gauge couplings \( g_1, g_2 \) based on the relationships \( m_W = \frac{v}{2}g_2 \) for the W boson mass and \( m_Z = \frac{v}{2}\sqrt{g_1^2 + g_2^2} \) for the Z boson mass. The bounds under the assumption of diminishing Higgs mass $m_H\rightarrow 0$ read, 
\begin{align}
   H H \rightarrow H H:\;\;\;\;B_\text{EW}^{(2)}=&\frac{g_1^2+3g_2^2}{4\pi^2\Lambda^2v^2},\;\;\;\;\;\;\;\;\;\;\; \;\;\;\;\;\;\;\;\;\;\;\;B_\text{GR}^{(2)}=\frac{8\sqrt{3}-125}{144\pi^2M_\text{Pl}^2v^2},\\   H \gamma \rightarrow H \gamma:\;\;\;\; B_\text{EW}^{(2)}=&\frac{2g_1^2 g_2^2}{\pi ^2 \Lambda ^2 v^2
		\left(g_1^2+g_2^2\right)},\;\;\;\;\;\;\;\;\;\;\nonumber\\B_\text{GR}^{(2)}=&\frac{8 \sqrt{3} \pi -125 }{288 \pi ^2 M_\text{Pl}^2 v^2}-\frac{7g_1^2 }{10 \pi ^2 M_\text{Pl}^2 v^2
   \left(g_1^2+g_2^2\right)}-\frac{11g_1^2 g_2^2}{720 \pi ^2 m_e^2 M_\text{Pl}^2
   \left(g_1^2+g_2^2\right)},\label{flip}\\   \gamma \gamma \rightarrow \gamma \gamma:\;\;\;\;B_\text{EW}^{(2)}=&\frac{32 g_1^4 g_2^2}{\pi ^2 \Lambda ^2 v^2
   \left(g_1^2+g_2^2\right)^2},\;\;\;\;\;\;\;\;\,B_\text{GR}^{(2)}=-\frac{g_1^2 \left(11 g_2^2 v^2+504 m_e^2\right)}{90 \pi ^2 m_e^2
   M_\text{Pl}^2 v^2 \left(g_1^2+g_2^2\right)}.
\end{align}
In the $H\gamma\rightarrow H\gamma$ process, it is essential to consider two types of loop corrections stemming from the 3-point vertices $HHh$ and $\gamma\gamma h$. The first type contributes to the amplitude proportional to the results in the $HH\rightarrow HH$ process and is represented by the first factor in equation \eq{flip}. The second type of correction results in the last two terms in the equation.

\section{Constraints on weak couplings}
\begin{figure}[h!]\centering
   \includegraphics[width=0.8\textwidth]{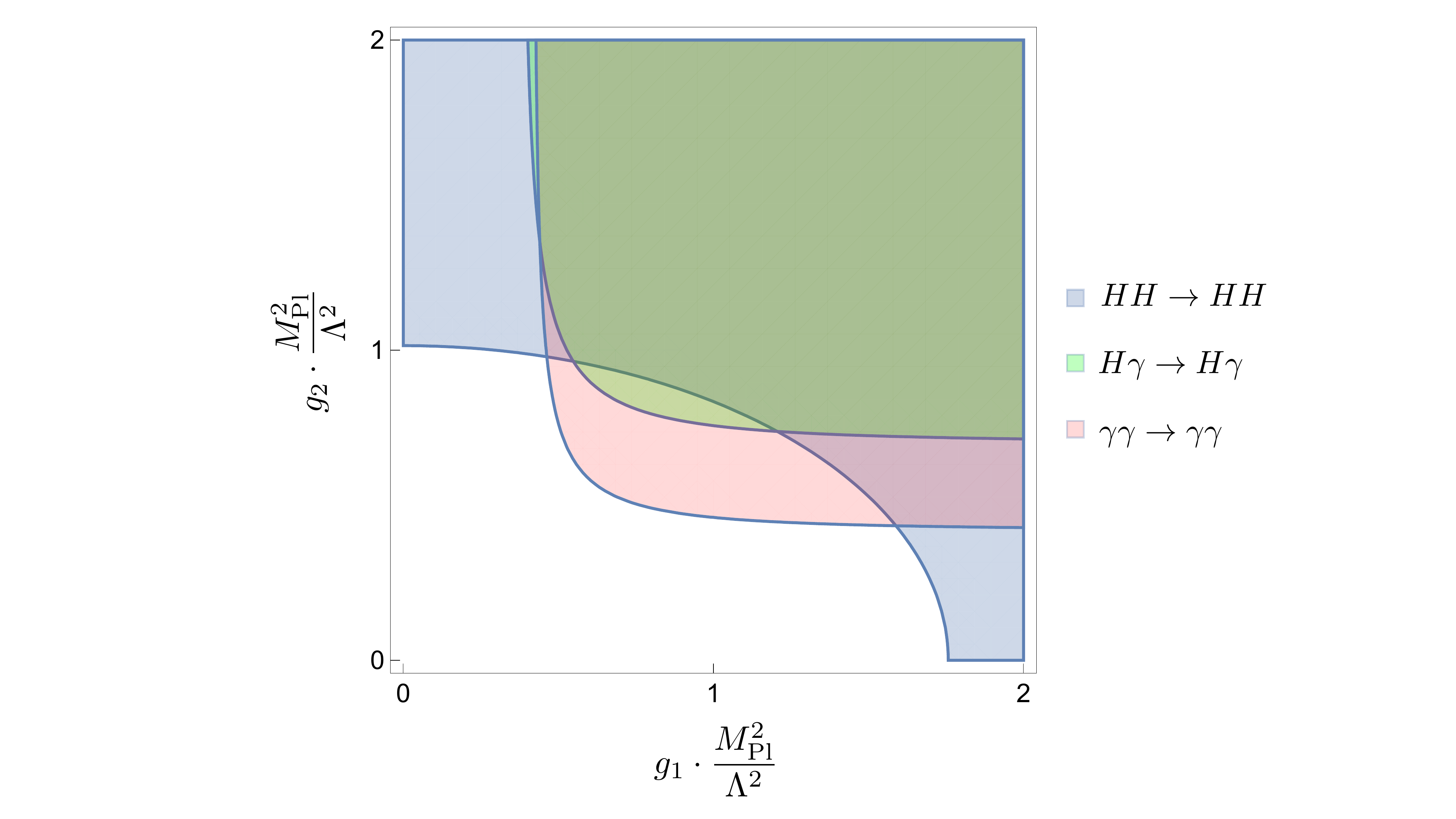}
   \caption{Constraints on Weak couplings}
   \label{Contraints}
\end{figure} 
In our initial setup, which assumes a zero Higgs mass, we further hypothesize a large fermion mass, denoted as $m_e\gg 1$.  Due to this assumption, factors of the order $\mathcal{O}(m_e^{-2})$ are suppressed. Consequently, from each $2\rightarrow 2$ process, we derive the following constraints for Weak couplings, depicted in \fig{Contraints}.
\begin{align}
	H H \rightarrow H H:\;\;\;\;\;\;\;\; &g_1^2+3g_2^2>\frac{125-8\sqrt{3}}{36}\frac{\Lambda^2}{M^2_\text{Pl}}.\\H \gamma \rightarrow H \gamma:\;\;\;\;\;\;\;\; &
\frac{125-8\sqrt{3}\pi}{576}\left(\frac{1}{g_1^2}+\frac{1}{g_2^2}\right)+\frac{7}{20}\frac{1}{g_2^2}<\frac{1}{\frac{\Lambda^2}{M^2_\text{Pl}}}.\\\gamma \gamma \rightarrow \gamma \gamma:\;\;\;\;\;\;\;\; &
	\frac{1}{g_1^2}+\frac{1}{g_2^2}<\frac{1}{\frac{11\frac{g_2v^2}{m_e^2}+504}{2880}\frac{\Lambda^2}{M^2_\text{Pl}}}\underset{\tiny{m_e\gg 1}}{\simeq}\frac{1}{\frac{7}{40}\frac{\Lambda^2}{M^2_\text{Pl}}},
\end{align}
\section{Conclusions}
In conclusion, we have found a possible connection between the WGC-like bounds and the positivity in the electroweak-like setup, extending upon the QED analysis. Our study revealed an interesting point: the constraints arising from $HH\rightarrow HH$ interactions disallow both couplings to be simultaneously small. Furthermore, the constraints from $\gamma H\rightarrow \gamma H$ and $\gamma\gamma\rightarrow\gamma\gamma$ interactions prevent either of them from being small on their own. Akin to the magnetic WGC, the constraints suggest that when the coupling constants $g_1, g_2$ are small, the effective theory breaks down at a relatively low scale.

\section*{Acknowledgement}
The results presented in this proceedings are based on an upcoming paper \cite{Ourpaper}. I would like to thank Katsuki Aoki, Toshifumi Noumi and Junsei Tokuda for their collaboration and comments on the present paper.

\bibliographystyle{JHEP}
\bibliography{Proceedings}







\end{document}